# Solution Of Wheeler-De Witt Equation, Potential Well And Tunnel Effect

Yong-Chang Huang [1,2,*]     Gang Weng [1]

1. Institute of Theoretical Physics, Beijing University of Technology, Beijing 100022, P. R. China
2. CCAST ( World Lab. ), P. O. Box 8730, Beijing, 100080, P. R. China

**Abstract**

This paper uses the relation of the cosmic scale factor and scalar field to solve Wheeler-DeWitt equation, gives the tunnel effect of the cosmic scale factor a and quantum potential well of scalar field, and makes it fit with the physics of cosmic quantum birth. By solving Wheeler-DeWitt equation we achieve a general probability distribution of the cosmic birth, and give the analysis of cosmic quantum birth.

## 1. Introduction

The standard cosmological theories successfully explain the evolvement of the universe from the big bang up to now. The theories can explain the phenomena such as the origin of the elements (for example, helium), the redshift of galaxy spectrum, 2.7K microwave background radiation, the galaxy count, the homogeneous property and isotropy in great scale and so on.

While there exists unfathomed difficult problem. Especially there exists the problem about the cosmic evolution details of the period from t =0 to $10^{-10}$ second. The inflationary cosmological model can solve the problem such as the magnetic monopole problem，the horizon difficulty，the problem of flat property.

The quantum cosmology gives a hope to solve the problem of the birth of the universe. The cosmic wave function describes the quantum state of the universe. The cosmic wave function satisfies the cosmological kinetic Wheeler-De Witt equation. By confirming boundary conditions, one can quantitatively study the problem about the birth of the universe. Wheeler-DeWitt equation is a Schrödinger equation of zero eignvalue [1].

---

* Corresponding author, email address: ychuang@bjut.edu.cn

Confirming the boundary conditions is to confirm the integral path of the ground state wave function, the path integral of the quantum mechanics is to sum the history.

The nowadays universe is isotropy in great scale, homogenous property and almost flat. The natures are that the universe has experienced the inflation of extreme high temperature and matter density [2]. The inflation makes temperature descend and the matter becomes sparse. After the big Bang the universe enters grand unified period, namely, unification period of the strong, weak and electromagnetic interactions. Because the CP breaking makes the universes have more baryon than anti-baryon and more positive matter than anti-matter, atomic nucleus are formed when temperature further declines, such as helium and deuteron. One of the missions of the quantum cosmology is to ensure the probability distribution of the early cosmic state. The early cosmic state can be related by scalar field $\Phi$ [3-6].

In chaotic inflationary theories [7], the probability of the cosmic quantum birth is marked by $\rho$. When $\Phi$ is the greatest, $\rho$ may tend to infinitude, the universe inflates extremely, the scalar field $\Phi$ can no longer roll down [8]. Therefore, the great field need be cut off, it is called the great field difficulty.

Wheeler-De Witt equation [9] beneath is established to describe the quantum birth of the universe

$$\frac{\partial^2 \Psi}{\partial a^2} - \frac{6}{k^2 a^2}\frac{\partial^2 \Psi}{\partial \Phi^2} - \frac{144\pi^4}{k^4}(k_c a^2 - \frac{k^2}{3}a^4 V(\Phi))\Psi = 0 \quad , \tag{1.1}$$

where $a$ is the cosmic scale factor, $\Phi$ the scalar field, $\Psi$ cosmic wave function and V($\Phi$) the scalar potential, $k_c$ may equal to one, k equals to the reciprocal of Plank mass $M_p$. In fact, Eq.(1.1) satisfies quantitative causal relation, i.e., which means that changes ( cause ) of some quantities in Eq.(1.1) must cause relative changes ( result ) of the other quantities in Eq.(1.1) so that the equation's right side keeps invariant. The cosmic naissance experienced an inflationary epoch and then enters the slow rolling dilation, it may be the gentle dilation or the frequent stir dilation.

Refs.[10,11] use "twice loosing shoe" to solve the Weeler-De Witt equation, and to study the tunneling effect of cosmic scale factor and the potential well effect of the scalar field $\Phi$, which can conquer the difficulty of the great field in the cosmic naissance, the potential well effect of $\Phi$ provides the truncation of the great field. If the early field $\Phi$ is not great, the universe maybe have the chaotic dilation.

The literature [7,11] firstly fixes $\Phi$, namely take $\frac{\partial \Psi}{\partial \Phi} = 0$, then it follows from (1.1) that

$$\frac{\partial^2 \Psi}{\partial a^2} = \frac{144\pi^4}{k^4}(k_c a^2 - \frac{k^2}{3}a^4 V(\Phi))\Psi \tag{1.2}$$

Therefore the tunneling probability can be obtained （$k_c=1$）[11] as follows

$$\rho_a = c_1 \exp(\frac{-24\pi^2}{k^4 V(\Phi)}) \tag{1.3}$$

If it takes

$$V(\Phi) = \frac{m^2 \Phi^2}{2} \tag{1.4}$$

when $\Phi_0 \to \infty$, at the beginning the universe inflates extremely. It has the great field difficulty [7,11].

Then the literatures [7,11] fix $a$, i.e., order $\frac{\partial \Psi}{\partial a} = 0$, it follows that

$$\frac{\partial^2 \Psi}{\partial \Phi^2} = 8\pi^4 a^6 (V(\Phi) - \frac{3}{a^2 k^2})\Psi \tag{1.5}$$

In succession it resolve the formula （1.5），and discuss the property of the resolution. It is obvious that such study is quite especial study.

This paper follows the researches of Refs.[6,7,10,11], generalizes their studies to more general situation, and investigates the cosmic birth. The concrete arrangement is: Sect. 2 is solution of the Wheeler-DeWitt equation for the cosmic evolution dynamics, Sect.3 researches the tunneling and potential well effects of cosmic quantum birth and evolution, the solution of cosmic quantum birth and evolution is discussed in different situations in Sect. 4, the last section is the summary and conclusion.

## 2. Solution of Wheeler-De Witt equation for the cosmic evolution dynamics

According to the standpoint of the inflationary cosmology, the universe originates from the decay of scalar field $\Phi$, so that the cosmic scalar factor depends on scalar field $\Phi$. So many documents such as document [1,7,10] adopts $a = \frac{\alpha}{V(\Phi)}$ ($\alpha$ is a parameter), when adopting the potential $V = \lambda_0 \Phi^2$ [7,11] ($\lambda_0$ is a parameter), the potential is mainly adopted in many inflationary cosmological models, thus we can acquire the function representation $a = \alpha_0 \Phi^{-2}$. It follows that $a = \alpha_0 \Phi^{-2} = \frac{\alpha_0 \lambda_0}{V}$, that is $\alpha_0 = \frac{\alpha}{\lambda_0}$ serves as the parameter. In this way, $a$ is guaranteed simultaneously to be invariant when $\Phi$ is transformed into $-\Phi$. And $a$ is largened when $\Phi$ is dwindled. As a consequence it follows that

$$\frac{\partial a}{\partial \Phi} = -2\alpha_0 \Phi^{-3} \qquad (2.1)$$

The cosmic kinetic Weeler—De Witt equation can be changed by using

$$\frac{1}{a^2}\frac{\partial^2}{\partial \Phi^2} = \frac{1}{a^2}(\frac{\partial a}{\partial \Phi})^2 \frac{\partial^2 \Psi}{\partial a^2} \qquad , \qquad (2.2)$$

where the exchangeability of $\frac{\partial}{\partial a}$ and $\frac{\partial}{\partial \Phi}$ is utilized. Then formula (1.1) can be changed into

$$\frac{\partial^2 \Psi}{\partial a^2} - \frac{6}{k^2} \frac{1}{a^2}(\frac{\partial a}{\partial \Phi})^2 \frac{\partial^2 \Psi}{\partial a^2} - \frac{144\pi^4}{k^4}（k_c a^2 - \frac{k^2}{3}a^4 V（\Phi））\Psi = 0 \quad (2.3)$$

Using $\Phi = (\frac{a}{\alpha_0})^{\frac{-1}{2}}$ we can obtain

$$\frac{6}{k^2 a^2}\frac{\partial^2 \Psi}{\partial \Phi^2} = \frac{24a}{\alpha_0 k^2}\frac{\partial^2 \Psi}{\partial a^2} \quad (2.4)$$

Thus there is

$$(1 - \frac{24a}{\alpha_0 k^2})\frac{d^2 \Psi}{da^2} = \frac{144\pi^4}{k^4}（k_c a^2 - \frac{k^2}{3}a^3 \alpha_0 \lambda_0）\Psi \quad (2.5)$$

Because there is a relation

$$\frac{1}{\Psi}\frac{d^2 \Psi}{da^2} = \frac{d}{da}(\frac{d\ell n \Psi}{da}) + (\frac{d\ell n \Psi}{da})^2 \quad (2.6)$$

then we can obtain

$$\ell n\Psi + ca + c_0 + (\ell n\Psi + c_1)^2 = G \quad (2.7)$$

where $c_0, c_1$ and c is the constants of integration, and there is

$$G = \iint \frac{144\pi^4 k^{-4}}{1 - \frac{24a}{\alpha_0 k^2}}（k_c a^2 - \frac{k^2}{3}a^3 \alpha_0 \lambda_0）\, da\, da \quad (2.8)$$

By carefully calculating Eq.(2.8), we can obtain

$$G(a) = \frac{-6\pi^4 \alpha_0}{k^2}\iint \frac{1}{a - \frac{k^2 \alpha_0}{24}}（k_c a^2 - \frac{k^2}{3}a^3 \alpha_0 \lambda_0）\, da\, da$$

$$= \frac{-6\pi^4 \alpha_0}{k^2}\{k_c [\frac{a^3}{6} + \frac{k^2 \alpha_0}{24}\frac{a^2}{2} + (\frac{k^2 \alpha_0}{24})^2 (a - \frac{k^2 \alpha_0}{24})\ell n\left|a - \frac{k^2 \alpha_0}{24}\right|$$

$$-(\frac{k^2 \alpha_0}{24})^2 a] - [\frac{a^4}{12} + (\frac{k^2 \alpha_0}{24})\frac{a^3}{6} + (\frac{k^2 \alpha_0}{24})^2 \frac{a^2}{2}$$

$$+ \ (\frac{k^2\alpha_0}{24})^3 (a - \frac{k^2\alpha_0}{24}) \ln\left|a - \frac{k^2\alpha_0}{24}\right| - (\frac{k^2\alpha_0}{24})^3 a \ ] + s_0 a + s_1 \} \qquad (2.9)$$

where $s_0$ and $s_1$ are the integration constants, the absolute value in Eq. (2.9) can guarantee that $a$ can take less value than $\frac{k^2\alpha_0}{24}$.

Defining $ln\Psi = f$, formula (2.7) can be rewrite as

$$f^2 + (2c_1 + 1)f + ca + c_0 - G = 0 \qquad (2.10)$$

$$f = \frac{-(2c_1+1) \pm \sqrt{(2c_1+1)^2 - 4(ca + c_0 - G)}}{2} \qquad (2.11)$$

So we can obtain

$$\Psi(a) = c_2 \exp\{\pm [G(a) - ca + (c_1 + \frac{1}{2})^2 - c_0]^{\frac{1}{2}}\} \qquad (2.12)$$

where $c_2 = \exp\{-(c_1 + \frac{1}{2})\}$. Because there is a relation $a = \alpha_0 \Phi^{-2}$, we have

$$\Psi(\Phi) = c_2 \exp\{\pm [G(\Phi) - c\alpha_0 \Phi^{-2} + (c_1 + \frac{1}{2})^2 - c_0]^{\frac{1}{2}}\} \qquad (2.13)$$

Substituting $a = \alpha_0 \Phi^{-2}$ into (2.9) we obtain

$$G(\Phi) = \frac{-6\pi^4 \alpha_0}{k^2} \{k_c [\frac{\alpha_0^3 \Phi^{-6}}{6} + \frac{k^2\alpha_0}{24} \frac{\alpha_0^2 \Phi^{-4}}{2} + (\frac{k^2\alpha_0}{24})^2 (\alpha_0 \Phi^{-2} - \frac{k^2\alpha_0}{24}) \ln\left|\alpha_0 \Phi^{-2} - \frac{k^2\alpha_0}{24}\right|$$

$$- (\frac{k^2\alpha_0}{24})^2 \alpha_0 \Phi^{-2}] - [\frac{\alpha_0^4 \Phi^{-8}}{12} + (\frac{k^2\alpha_0}{24}) \frac{\alpha_0^3 \Phi^{-6}}{6} + (\frac{k^2\alpha_0}{24})^2 \frac{\alpha_0^2 \Phi^{-4}}{2} +$$

$$(\frac{k^2\alpha_0}{24})^3 (\alpha_0 \Phi^{-2} - \frac{k^2\alpha_0}{24}) \ln\left|\alpha_0 \Phi^{-2} - \frac{k^2\alpha_0}{24}\right| - (\frac{k^2\alpha_0}{24})^3 \alpha_0 \Phi^{-2}] + s_0 \alpha_0 \Phi^{-2} + s_1\}$$
$$(2.14)$$

So formula (2.12) can be concretely rewritten as

$$\Psi(a) = c_2 \exp\{\pm \{\frac{-6\pi^4 \alpha_0}{k^2} \{k_c [\frac{a^3}{6} + \frac{k^2\alpha_0}{24} \frac{a^2}{2} + (\frac{k^2\alpha_0}{24})^2 (a - \frac{k^2\alpha_0}{24}) \ln\left|a - \frac{k^2\alpha_0}{24}\right|$$

$$- (\frac{k^2\alpha_0}{24})^2 a] - \frac{k^2}{3}\alpha_0\lambda_0 [\frac{a^4}{12} + (\frac{k^2\alpha_0}{24}) \frac{a^3}{6} + (\frac{k^2\alpha_0}{24})^2 \frac{a^2}{2} + (\frac{k^2\alpha_0}{24})^3 (a - \frac{k^2\alpha_0}{24}) \ln\left|a - \frac{k^2\alpha_0}{24}\right|$$

$$- (\frac{k^2\alpha_0}{24})^3 \ a \ ] + s_0 a + s_1 \} - c a + (c_1 + \frac{1}{2})^2 - c_0\}^{\frac{1}{2}}\} \qquad (2.15)$$

We further discuss the tunneling and potential well effects of the cosmic generation as follows.

## 3. The tunneling and potential well of the cosmic quantum generation and evolution

From Eq.(2.5) we can obtain

$$\frac{d^2\Psi}{da^2} = \frac{144\pi^4}{k^4} (k_c a^2 - \frac{k^2}{3} a^3 \alpha_0 \lambda_0)(1 - \frac{24a}{\alpha_0 k^2})^{-1} \Psi \qquad (3.1)$$

By using Schrödinger equation

$$H\Psi = E\Psi, \qquad H = -\frac{\hbar^2}{2m}\frac{\partial^2}{\partial x^2} + U(x) \qquad (3.2)$$

It follows that [11]

$$\frac{d^2}{dx^2}\Psi = \frac{2m}{\hbar^2}(U(x) - E)\Psi \qquad (3.3)$$

Then we can obtain effective potential

$$U_1(a) = \frac{144\pi^4}{k^4}(k_c a^2 - \frac{k^2}{3} a^3 \alpha_0 \lambda_0)(1 - \frac{24a}{\alpha_0 k^2})^{-1}, \quad E_a = 0 \qquad (3.4)$$

When $a_m = \frac{\alpha_0 k^2}{24}$, $U_1(a)$ has the maximum value so that we find that $U_1(a)$ is a infinitude deep potential well. This is the conclusion that people did not obtain in the past, for instance, literatures [1,7,10,11] didn't have this conclusion. When $a$ change from $a = 0$ to $a_0$ and $a_0 > a_m$, $U(a)$ alters into $U(a_0) = 0$ and there is tunneling through barrier.

By using $a = \alpha_0 \Phi^{-2}$ and $da = -2\alpha_0 \Phi^{-3} d\Phi$ we obtain

$$\frac{d^2\Psi}{d\Phi^2} = \frac{576\pi^4}{k^4}(k_c \alpha_0^2 \Phi^{-4} - \frac{k^2}{3}\alpha_0^4 \lambda_0 \Phi^{-6})(1 - \frac{24}{k^2}\Phi^{-2})^{-1}\alpha_0^2 \Phi^{-6}\Psi \qquad (3.5)$$

Similarly we can obtain the effective potential about the scalar field $\Phi$ as follow

$$U_2(\Phi) = \frac{576\pi^4}{k^4}(k_c \alpha_0^2 - \frac{k^2}{3}\alpha_0^4 \lambda_0 \Phi^{-2})(1 - \frac{24}{k^2}\Phi^{-2})^{-1}\alpha_0^2 \Phi^{-10} \quad (3.6)$$

It is obvious that $U_2(\Phi)$ tends to infinitude when $\Phi = 0$ and $\Phi^2 = \frac{k^2}{24}$. Therefore there is infinitude deep potential well on the situation of $\Phi \in (0, \frac{\sqrt{24}}{k})$. When $a$ changes from $a = 0$ to $a_0$, $U(a)$ alters into $U(a_0) = 0$ and there is tunneling through barrier. Then $\Phi = \infty$ is changed into $\Phi_0$ and $U(\Phi)$ is changed into $U(\Phi_0) = 0$. namely, there is quantum well effect. By means of $a = \alpha_0 \Phi^{-2}$, we obtain the conclusion that $\Phi$ decreases when $a$ aggrandizes.

## 4. Discussion of the solution of the cosmic quantum generation

Because Eq.(2.15) is a general wave function about the cosmic quantum generation, we need discuss the problem about sign $\pm$ in the exponential wave function. In the location where there is $a \to \varepsilon$ from zero, known as the infinitesimal value of $a$ when the universe just forms the micronucleus, Eq.(2.15) can be left the term that have the main contribution for the wave function, that is, we can remove the high order term of $a$. Then we can obtain

$$\Psi(a) = c_2 \exp\{\pm\{\frac{-6\pi^4 \alpha_0}{k^2}\{k_c [(\frac{k^2\alpha_0}{24})^2 (a - \frac{k^2\alpha_0}{24})\ell n|a - \frac{k^2\alpha_0}{24}| - (\frac{k^2\alpha_0}{24})^2 a]$$

$$-\frac{k^2}{3}\alpha_0 \lambda_0 [(\frac{k^2\alpha_0}{24})^3 (a - \frac{k^2\alpha_0}{24})\ell n|a - \frac{k^2\alpha_0}{24}| - (\frac{k^2\alpha_0}{24})^3 a] + s_0 a + s_1\} - c a$$

$$+ (c_1 + \frac{1}{2})^2 - c_0\}^{\frac{1}{2}}\} \quad (4.1)$$

When there is

$$\frac{-6\pi^4 \alpha_0}{k^2}\{k_c [(\frac{k^2\alpha_0}{24})^2 (a - \frac{k^2\alpha_0}{24})\ell n|a - \frac{k^2\alpha_0}{24}| - (\frac{k^2\alpha_0}{24})^2 a] - \frac{k^2}{3}\alpha_0 \lambda_0 [(\frac{k^2\alpha_0}{24})^3$$

$$\bullet\ (a-\frac{k^2\alpha_0}{24})\ln\left|a-\frac{k^2\alpha_0}{24}\right|-(\frac{k^2\alpha_0}{24})^3 a]+s_0 a+s_1\}-c\,a+(c_1+\frac{1}{2})^2-c_0=-L<0$$

It follows that

$$\Psi(a)=c_2\exp(\pm iL^{1/2}) \qquad (4.2)$$

where L is the positive real function. In the location $a=\varepsilon$, the solution varies by means of the form $\exp(\pm i L^{1/2})$. The cosmic quantum generation probability is $\rho=c_2^2$.

At the great value $a$, $a^4$ predominates. Then we can obtain

$$\Psi(a)=c_2\exp\{\pm\{\frac{\pi^4 a^4 \alpha_0^2 \lambda_0}{6}\}^{\frac{1}{2}}\} \qquad (4.3)$$

When the exponent of $\Psi(a)$ takes the positive sign, $a$ increases, $\Psi(a)$ exponentially increases, then the probability exponentially increases such that the universe aggrandizes to inflate [7,10].

Due to $a=\alpha_0\Phi^{-2}$, its corresponding wave function about $\Phi$ is as follow: at this point, there is $\Phi\to\varepsilon_1$ and $\varepsilon_1$ is the minimum such that $\Phi$ can be adopted as

$$\Psi(\Phi)=c_2\exp\{[\frac{\pi^4\alpha_0^6\lambda_0}{6\Phi^8}]^{\frac{1}{2}}\} \qquad (4.4)$$

When the exponent of $\Psi(\Phi)$ still aggrandizes, that the matter field $\Phi$ decays causes the universe to inflate. Therefore, the two kinds of discussions above are consistent, and our researches don't have the great field difficulty in the literature [7,11]. The general probability of the cosmic quantum generation is as follow

$$\rho(a)=c_2^2\exp\{2\{\frac{-6\pi^4\alpha_0}{k^2}\{k_c[\frac{a^3}{6}+\frac{k^2\alpha_0}{24}\frac{a^2}{2}+(\frac{k^2\alpha_0}{24})^2(a-\frac{k^2\alpha_0}{24})\ln\left|a-\frac{k^2\alpha_0}{24}\right|$$

$$-(\frac{k^2\alpha_0}{24})^2 a]-\frac{k^2}{3}\alpha_0\lambda_0[\frac{a^4}{12}+(\frac{k^2\alpha_0}{24})\frac{a^3}{6}+(\frac{k^2\alpha_0}{24})^2\frac{a^2}{2}+(\frac{k^2\alpha_0}{24})^3(a-\frac{k^2\alpha_0}{24})$$

$$\bullet\ \ln\left|a-\frac{k^2\alpha_0}{24}\right|-(\frac{k^2\alpha_0}{24})^3 a]+s_1 a+s_2\}-c\,a+(c_1+\frac{1}{2})^2-c_0\}^{\frac{1}{2}}\} \qquad (4.5)$$

Therefore, using $a=\alpha_0\Phi^{-2}$ and Eq.(2.15), we can obtain the general concrete representation of $\Psi(\Phi)$

$$\Psi(\Phi) = c_2 \exp\{\{\frac{-6\pi^4 \alpha_0}{k^2} \{k_c [\frac{\alpha_0^3 \Phi^{-6}}{6} + \frac{k^2 \alpha_0}{24}\frac{\alpha_0^2 \Phi^{-4}}{2} + (\frac{k^2 \alpha_0}{24})^2 (\alpha_0 \Phi^{-2} - \frac{k^2 \alpha_0}{24})$$

$$\cdot \ln\left|\alpha_0 \Phi^{-2} - \frac{k^2 \alpha_0}{24}\right| - (\frac{k^2 \alpha_0}{24})^2 \alpha_0 \Phi^{-2}] - \frac{k^2}{3}\alpha_0 \lambda_0 [\frac{\alpha_0^4 \Phi^{-8}}{12} + (\frac{k^2 \alpha_0}{24}) \frac{\alpha_0^3 \Phi^{-6}}{6}$$

$$+ (\frac{k^2 \alpha_0}{24})^2 \frac{\alpha_0^2 \Phi^{-4}}{2} + (\frac{k^2 \alpha_0}{24})^3 (\alpha_0 \Phi^{-2} - \frac{k^2 \alpha_0}{24}) \ln\left|\alpha_0 \Phi^{-2} - \frac{k^2 \alpha_0}{24}\right| - (\frac{k^2 \alpha_0}{24})^3 \alpha_0 \Phi^{-2}] +$$

$$s_0 \alpha_0 \Phi^{-2} + s_1\} - c\,\alpha_0 \Phi^{-2} + (c_1 + \frac{1}{2})^2 - c_0\}^{\frac{1}{2}}\} \tag{4.6}$$

The general probability of the cosmic quantum birth is

$$\rho(\Phi) = c_2 \exp\{2\{\frac{-6\pi^4 \alpha_0}{k^2} \{k_c [\frac{\alpha_0^3 \Phi^{-6}}{6} + \frac{k^2 \alpha_0}{24}\frac{\alpha_0^2 \Phi^{-4}}{2} + (\frac{k^2 \alpha_0}{24})^2 (\alpha_0 \Phi^{-2} - \frac{k^2 \alpha_0}{24})$$

$$\cdot \ln\left|\alpha_0 \Phi^{-2} - \frac{k^2 \alpha_0}{24}\right| - (\frac{k^2 \alpha_0}{24})^2 \alpha_0 \Phi^{-2}] - \frac{k^2}{3}\alpha_0 \lambda_0 [\frac{\alpha_0^4 \Phi^{-8}}{12} + (\frac{k^2 \alpha_0}{24}) \frac{\alpha_0^3 \Phi^{-6}}{6}$$

$$+ (\frac{k^2 \alpha_0}{24})^2 \frac{\alpha_0^2 \Phi^{-4}}{2} + (\frac{k^2 \alpha_0}{24})^3 (\alpha_0 \Phi^{-2} - \frac{k^2 \alpha_0}{24}) \ln\left|\alpha_0 \Phi^{-2} - \frac{k^2 \alpha_0}{24}\right|$$

$$-(\frac{k^2 \alpha_0}{24})^3 \alpha_0 \Phi^{-2}] + s_0 \alpha_0 \Phi^{-2} + s_1\} - c\,\alpha_0 \Phi^{-2} + (c_1 + \frac{1}{2})^2 - c_0\}^{\frac{1}{2}}\} \tag{4.7}$$

It is obvious that our method is more general and differs from the method of "twice loose shoe" in the literatures [7, 11]. Because the universe originates from the decay of scalar field $\Phi$, the cosmic radius $a$ depends upon the scalar field $\Phi$.

In words, the universe goes through extreme inflation, or chaotic inflation, then enters the slow rolling dilatation until the end, finally there is only minor quantum stir left, the universe further enters the standard describing stage by means of cosmological models of the big bang. About researches on the generalization of $a = \alpha_0 \Phi^{-2}$ to general expression $a = a(\Phi)$ and applications etc in quantum cosmology will be written in the other papers.

## 5. Summary and conclusion

This paper solves Wheeler-De Witt equation of describing the cosmic quantum birth and the inflation that the universe has experienced, and overcomes the scarcity of the method of "twice loose shoe". Because the method of "twice loose shoe" [7,11] seeks to use, in turn, the cosmic scale factor $a$ ( in the same time, demand $\frac{\partial \Psi}{\partial \Phi}= 0$ ) and the scalar field $\Phi$ ( in the same time, demand $\frac{\partial \Psi}{\partial a}= 0$ ) to solve the Wheeler-De Witt equation, and expects to conquer the great field difficulty that is caused by only considering the tunneling effect of the cosmic quantum generation, namely, using the potential well effect of $\Phi$ to offer the truncation of the great field [11].

This paper uses the relation of the cosmic scale factor and scalar field to solve Wheeler-DeWitt equation, gives the tunnel effect of the cosmic scale factor a and quantum potential well of scalar field, and makes it match the physics of cosmic quantum birth. By solving Wheeler-De Witt equation we achieve the probability distribution of the cosmic generation, and give the analysis of cosmic quantum birth relative to Wheeler-De Witt equation.

Because of $a=\alpha_0 \Phi^{-2}$, when $a$ aggrandizes, $\Phi$ dwindles, after the infinitesimal universe born, the universe experienced inflation, or critical chaotic inflation, and then enters into the slow rolling stage of the dilation, it ends in the small quantum stir, then it enters the standard describing stage by means of the big bang cosmology.

One ( Y. C. Huang ) of the authors is grateful for Prof. D. H. Zhang for useful discussion on some problems relative to this paper in different periods.